\documentclass[preprint,prb,superscriptaddress,amssymb,floatfix,citeautoscript]{revtex4-1}

\usepackage{subfigure}
\usepackage{graphicx}  
\usepackage{dcolumn}   
\usepackage{bm}        
\usepackage{amssymb}   

\newcommand{\w}{\omega}

\begin{document}

\title{Importance of frequency-dependent grain boundary scattering in nanocrystalline silicon and silicon-germanium thermoelectrics}

\author{Chengyun Hua}
\author{Austin.\ J. Minnich}
\email{aminnich@caltech.edu}
\affiliation{Division of Engineering and Applied Science,\\
California Institute of Technology, Pasadena, CA 91125}

\begin{abstract}

Nanocrystalline silicon and silicon-germanium alloys are promising thermoelectric materials that have achieved substantially improved figure of merits compared to their bulk counterparts. This enhancement is typically attributed to a reduction in lattice thermal conductivity by phonon scattering at grain boundaries. However, further improvements are difficult to achieve because grain boundary scattering is poorly understood, with recent experimental observations suggesting that the phonon transmissivity may depend on phonon frequency rather than being constant as in the commonly used gray model. Here, we examine the impact of frequency-dependent grain boundary scattering in nanocrystalline silicon and silicon-germanium alloys in a realistic 3D geometry using frequency-dependent variance-reduced Monte Carlo simulations. We find that the grain boundary may not be as effective as predicted by the gray model in scattering certain phonons, with a substantial amount of heat being carried by low frequency phonons with mean free paths longer than the grain size. Our result will help guide the design of more efficient thermoelectrics. 

\end{abstract}

\maketitle    

\section{Introduction}

Thermoelectric (TE) materials, which can convert heat directly to electricity, are of considerable interest for applications such as waste heat recovery due to their silent operation, reliability and lack of working fluid.\cite{Tian2013,Heremans2013,Yee2013} The performance of TE devices is characterized by the thermoelectric figure of merit $zT = S^2\sigma T/(k_e+k_{ph})$, where $S$ is the Seebeck coefficient, $\sigma$ is the electrical conductivity, $T$ is the absolute temperature at which the properties are measured and $k_e$ and $k_{ph}$ are the corresponding thermal conductivities for electrons and phonons.\cite{Gangbook} To achieve comparable efficiency to that of mechanical cycles, $zT$ should be greater than 2, but commonly used bulk TE materials such as $\text{Bi}_2\text{Te}_3$ typically have $zT < 1$.

Recently, nanocrystalline materials, which are polycrystalline materials with nanoscale grain sizes, have been demonstrated as efficient thermoelectric by significantly reducing the phonon thermal conductivity by strong phonon grain boundary scattering.\cite{Bux:2009,SciencePaper,SiGeNanoLett,XWWang2008,Mehta:2012,Biswas2011,Biswas2012} Many of these materials show great promise for scalable manufacturing. In particular, the nanostructuring approach has been successful for silicon and silicon-germanium alloys created by ball-milling and hot pressing. SiGe has long been used for space power generation\cite{Dismukes1964,RoweSiGe81} and the thermoelectric properties of SiGe nanocomposites with improved properties over those of the bulk were recently reported.\cite{RoweSiGe81,SiGeNanoLett,XWWang2008,Ren2009} Substantial improvements in nanostructured bulk Si, which does not require expensive and rare Ge, were also reported.\cite{Bux:2009} While these gains are significant, further improvements are difficult to achieve due to the lack of understanding of grain boundary scattering, which plays a critical role in scattering phonons in nanocrystalline materials.

Phonon transport in nanostructured Si/SiGe has been recently studied using molecular dynamics\cite{Cruz:2013,Colombo:2014,Keblinski:2006,Schelling:2002,Kimmer2007}, atomistic Green's functions\cite{Freund:2009,Ronggui:2012,Tian2012} and phonon-Boltzmann-equation-based simulations.  \cite{Hao:2010,Hao:2012b,Hao:2012a,Ronggui:2010,Jeng:2008,minnich:155327,Mingo:2010} Many studies assumed that the transmissivity of phonons across the boundary to be constant in a gray model. However, atomistic calculations\cite{Schelling:2002,Kimmer2007,Ronggui:2012,Tian2012} reported that transmissivity depends on phonon frequency. Kimmer \emph{et al.}\cite{Kimmer2007} found that transmission through the high-energy grain boundary in silicon is a function of the phonon frequency using molecular-dynamics simulations. Li and Yang\cite{Ronggui:2012} used atomistic Green's functions to study the effect of lattice mismatch on phonon transmission across interfaces and showed that transmissivity decreases as phonon frequency increases. These predictions are supported by a recent experimental work by Wang \emph{et al}\cite{Wang:2011a}, who reported temperature ($T$) dependent measurement of thermal conductivities of nanocrystalline silicon. While a $T^3$ dependence of thermal conductivity is expected based on the gray model, they instead observed a $T^2$ dependence, implying that transmissivity increases with decreasing phonon frequency. This result suggests that lower frequency modes are less scattered by interfaces compared to higher frequency modes. A frequency-dependent transmissivity would have an important effect on the distribution of heat among the phonon spectrum and hence on strategies to further reduce the thermal conductivity of nanocrystalline Si and SiGe. However, phonon transport in nanocrystalline thermoelectrics with frequency dependent grain boundary scattering has not yet been systematically explored.

Here we examine the effects of frequency-dependent grain boundary scattering on the thermal conductivity of nanocrystalline Si and SiGe alloys using efficient variance-reduced Monte Carlo (MC) simulations. This novel computational method enables the simulation of phonon transport in the full 3D geometry of the crystal with orders of magnitude reduced computational cost compared to previous numerical methods while rigorously including the frequency-dependence of phonon properties. Moreover, the simulations allow us to examine in detail how the frequency-dependent grain boundary scattering modifies the distribution of heat among the thermal phonon spectrum, providing an important guide to further improving the thermoelectric efficiency of nanocrystalline Si and SiGe.

\section{Modeling}

We seek to simulate thermal phonon transport in a polycrystalline domain with grain sizes of tens to hundreds of nanometers. To study heat transfer in this mesoscopic structure, we need to solve the frequency-dependent Boltzmann transport equation (BTE) under the relaxation time approximation. The energy-based BTE is given by:\cite{Majumdar:1993,Peraud:2011}
\begin{equation} \label{BTE_energy}
\frac{\partial e^d_{\omega}}{\partial t}+\mathbf{v}_g \cdot \nabla e^d_{\omega} = \frac{(e^{loc}_{\omega}-e^{eq}_{\omega})-e^d_{\omega}}{\tau(\omega,p,T)}
\end{equation}
where $e^d_{\omega}=\hbar\omega(f-f^{eq}_{T_{eq}})$ is the desired deviational distribution function, $f^{eq}_{T_{eq}}=[exp(\hbar\omega/k_bT_{eq})-1]^{-1}$ is a Bose-Einstein distribution at the control temperature $T_{eq}$, $\mathbf{v}_g(\omega,p)$ is the phonon group velocity and $\tau(\omega,p,T)$ is the phonon relaxation time. Here, $\hbar$ is the reduced Planck constant, $k_b$ is the Bolzmann constant, $T$ is the temperature, $\omega$ is the phonon frequency and p is the phonon polarization.

The BTE is very difficult to solve numerically even in simple geometries due to the high dimensionality of the distribution function. Various numerical techniques have been developed to solve the BTE, including discrete ordinates\cite{Majumdar:1993}, a finite volume approach\cite{Amon2004}, Monte Carlo simulations\cite{Hao:2010,Jeng:2008} and a mean free path (MFP) sampling algorithm.\cite{McGaughey2012} However, these methods are either computationally expensive or make simplifying assumptions that may not be accurate. Another approach to incorporate boundary scattering is to model the interface scattering using a phenomenological volumetric scattering rate $\tau^{-1} = v/L$.\cite{minnich:155327} While this approach has minimal computational requirements and provides useful insights, it lacks predictive power because the value of the geometrical length $L$ depends on frequency in the non-gray model and is not known in advance. 

In this work, we overcome these computational challenges using a novel variance-reduced, frequency-dependent Monte Carlo technique recently introduced by Peraud and Hadjiconstantinou. \cite{Peraud:2011,Peraud:2012,PeraudAnnRevHeatTransfer,PeraudMechEngRev,PeraudUnplished} Specifically, we solve the adjoint BTE using a linearized, energy-based deviational MC algorithm. This algorithm enables us to solve the frequency-dependent BTE in a large, complex geometry with minimal memory requirements and substantially reduced computational cost compared to traditional MC algorithms.

The algorithm we have chosen incorporates several key improvements over other numerical methods that we describe in turn. First, we use a deviational MC algorithm based on Eq.~(\ref{BTE_energy}) rather than traditional MC method.\cite{Peraud:2011} Deviational MC techniques achieve variance-reduction by recognizing that many of the collisions performed in traditional MC simulations serve only to stochastically compute the Bose-Einstein equilibrium distribution that is already known analytically. The variance of the calculation can be dramatically reduced by replacing the stochastic calculation of this known distribution with an analytical expression.

Second, for problems exhibiting sufficiently small temperature differences, the collision operator in Eq.~(\ref{BTE_energy}) can be linearized. As a result, properties for scattered particles can be drawn from a distribution that is independent of the local temperature, allowing each particle to be simulated independently and thereby eliminating the need for discretization in space and time.\cite{Peraud:2012} In addition to further reducing computational cost, this method dramatically reduces the memory requirements of MC simulations because all the phonon properties do not need to be stored simultaneously.

Finally, we solve the adjoint BTE rather than the traditional BTE using a recent approach introduced by Peraud \emph{et al.}\cite{PeraudAnnRevHeatTransfer,PeraudMechEngRev,PeraudUnplished} One drawback of solving the traditional BTE is that the frequencies of the phonon are drawn from a distribution weighted by the density of states of each phonon mode. As a result, even if modes with small density of states contribute substantially to heat conduction, as occurs in silicon, these phonons are unlikely to be sampled, leading to large noise at certain frequencies. Peraud \emph{et al.} recently showed that this limitation can be overcome using an adjoint approach, in which phonons are emitted uniformly over the frequency spectrum and subsequently weighted by the correct distribution to provide a variance-reduced estimate of the heat flux.

The full details of the computation are discussed by Peraud \emph{et al.} and will only be briefly reviewed here. The computational domain is a 3D cube with square grains as illustrated in the inset of Fig.~\ref{fig:Validation}(b). The domain bisects the grain in each dimension, placing the corner of the grains in the center of the domain. We impose a periodic heat flux boundary conditions along the direction of temperature gradient to model an infinitely repeating structure.\cite{Hao:2009} The other two directions have a specular boundary condition imposed by symmetry. These boundary conditions allow the thermal properties of an infinite polycrystal to be calculated using only one period. 

We impose a linearly varying equilibrium temperature $T_{eq}(x)$ that allows the control temperature to follow the physical temperature more closely.\cite{Peraud:2012} This variation in $T_{eq}$ can be implemented as a uniform volumetric source of deviational phonon bundles, each representing a fixed amount of deviational energy. The frequencies of the phonon bundles are drawn from a uniform distribution according to the adjoint BTE. The algorithm then proceeds by stochastically simulating the advection, scattering and sampling of phonon bundles sequentially and completely independently exactly as described by Peraud \emph{et al}.\cite{Peraud:2012} Finally, the calculated spectral heat flux is weighted by the appropriate distribution to obtain the final heat flux.\cite{PeraudAnnRevHeatTransfer}

In this work, we consider Si and $\text{Si}_{1-x}\text{Ge}_{x}\ (x < 0.5)$ with Ge modeled as a mass defect scattering mechanism in silicon. We model silicon using the dispersion along the [100] direction and assume that the crystal is isotropic.\cite{Minnich:2011} Optical phonons are neglected due to their small contributions to heat transfer in silicon.\cite{Hao:2010} The acoustic phonon relaxation times are given by\cite{Minnich:2011,Holland,ase}:
\begin{eqnarray}
\tau_{L}^{-1}(\omega) &=& 2\times10^{-19} \times \w^{2} T^{1.49} \exp(-80/T), \label{Eq:Tau_pp_Long} \\
\tau_{T}^{-1}(\omega) &=& 1.2\times10^{-19} \times\w^{2} T^{1.65} \exp(-80/T), \label{Eq:Tau_pp_trans}\\
\tau_{MD}^{-1}(\omega) &=& A \w^{4}, \label{Eq:Tau_md} 
\end{eqnarray}
where $A$ is a constant that equals $3\times10^{-45}$ for bulk silicon\cite{Klemens:1955}. 

For electron-phonon scattering, the relaxation time is given as\cite{Ziman1960}
\begin{equation}\label{eq:tau_ep}
\tau^{-1}_{ep}(\omega) = \frac{E_d^2 m^{*3}v_s}{4\pi\hbar^4\rho}\frac{k_BT}{\frac{1}{2}m^*v^2_s}\left\{\frac{\hbar\omega}{k_BT}-\text{ln}\frac{1+\exp\left[\frac{\frac{1}{2}m^*v_s^2-E_f}{k_BT}+\frac{\hbar^2\omega^2}{8m^*v^2_sk_BT}+\frac{\hbar\omega}{2k_BT}\right]}{1+\exp\left[\frac{\frac{1}{2}m^*v_s^2-E_f}{k_BT}+\frac{\hbar^2\omega^2}{8m^*v^2_sk_BT}-\frac{\hbar\omega}{2k_BT}\right]}\right\}
\end{equation}
where $E_d$, $m^*$, $\rho$, $v_s$ and $E_f$ represent acoustic deformation potential, density of states effective mass, density, averaged phonon group velocity and Fermi level, respectively. The values of $m^*$, $v_s$, $\rho$ and $E_f$ are obtained from the literature, which are $m^* = 1.4 m_0$, $v_s = 6500\ \text{m/s}$, $\rho = 2500\ \text{kg/m}^2$, and $E_f = 0.05\ \text{eV}$ corresponding to a doping concentration of approximately $10^{20}\ \text{cm}^{-3}$ in n-type Si.

For mass defect scattering in SiGe, we take A to be $3\times10^{-42}$, which corresponds to x = 0.25 in $\text{Si}_{1-x}\text{Ge}_{x}$ according to the formula given by Klemens.\cite{Klemens:1955} We then adjust the deformation potential $E_d$ until the thermal conductivity matches experimentally reported values of bulk SiGe,\cite{Zebarjadi:2011b} yielding $E_d = 2.5\ \text{eV}$, in reasonable agreement with the value reported by Holland.\cite{Holland} For doped Si, we change A to be $3\times10^{-43}$, which corresponds to a doping concentration of $10^{20}\ \text{cm}^{-3}$.\cite{Hao:2010} In the next section, we show that the results of this work are not sensitive to the exact values of these fitting parameters.

The critical parameter in our model is the phonon transmissivity across the grain boundaries. Many previous works assumed a gray model, but this model is not consistent with the measurements of Wang \emph{et al.} We use a frequency-dependent transmissivity $t(\omega)$ proposed by Wang \emph{et al.} in the form of\cite{Wang:2011a}
\begin{equation}\label{Eq:Transmissivity}
t(\omega)=\frac{1}{\gamma\omega/\omega_{max}+1}, 
\end{equation}
where $\gamma$ is a fitting parameter and $\omega_{max}$ is the maximum phonon frequency. This model is qualitatively consistent with atomistic Green's function calculations.\cite{Freund:2009,Ronggui:2012} Note that this condition implies that the transmissivity approaches one as frequency goes to zero, which is physically consistent with the expectation that long wavelength phonons are unaffected by atomistic disorder at a grain boundary.

We implement the condition with the following procedure. If a particle encounters the interface, its transmissivity is calculated using the phonon's frequency. If a random number is less than the calculated transmissivity, then the phonon transmits through the interface; otherwise, the phonon is reflected. Following Wang \emph{et al.}, we use Ziman's specularity parameter to determine whether the scattering is specular or diffuse:
\begin{equation}\label{Eq:specularity}
p(\omega,p) = \exp\left[-\frac{16\pi^2\eta^2}{\lambda^2(\omega,p)}\right],
\end{equation}
where $\lambda(\omega,p)$ is the phonon wavelength and $\eta$ is the rms surface roughness. If a randomly drawn number is less than $p$, the phonon experiences specular scattering, in which case the phonon's velocity vector is known from momentum conservation. Otherwise, the phonon is scattered diffusely. In the latter case, the phonon's new velocity vector is randomized according to the half-sphere defined by the normal vector of the interface.

\section{Results and discussion}

\begin{figure*}[t!]
\centering
\includegraphics[scale=0.5]{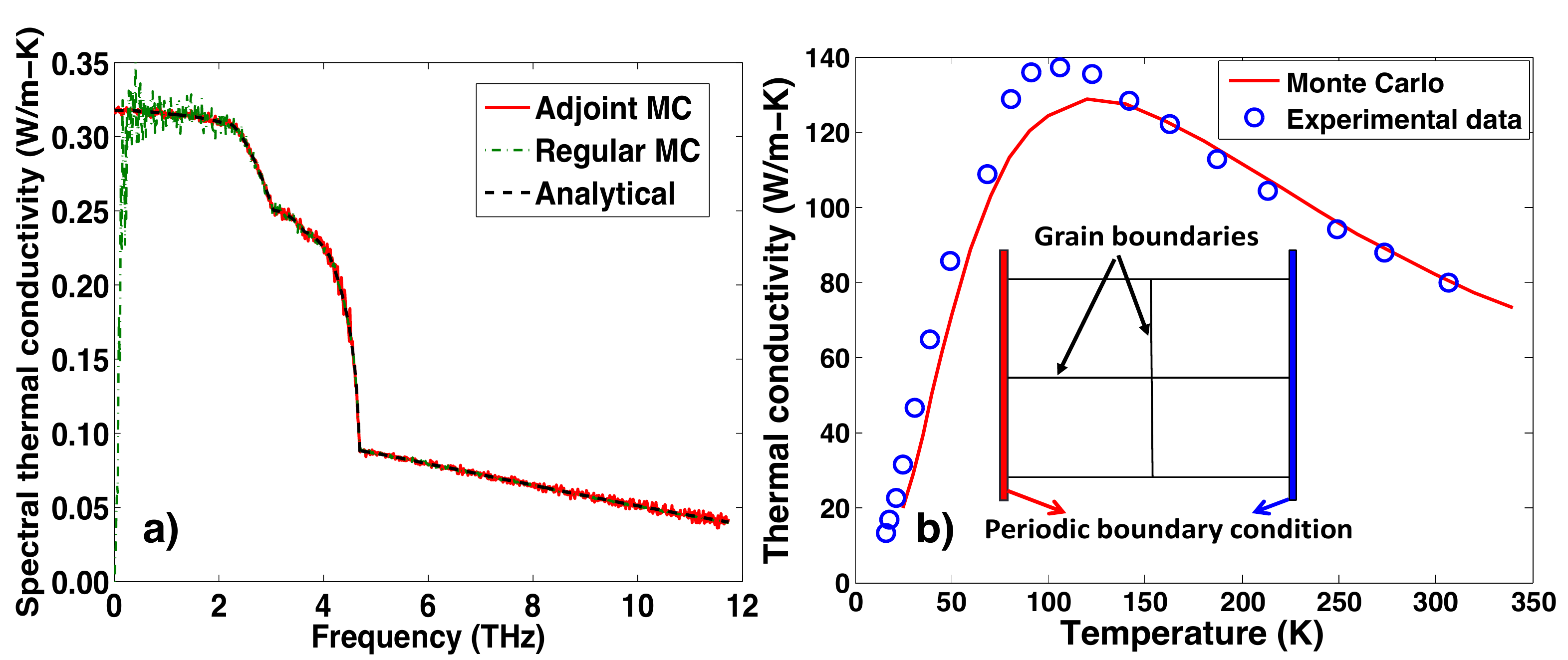}
\caption{(a) Spectral thermal conductivity plotted as a function of phonon frequency for bulk silicon at 300 K: adjoint MC method (solid line); original MC algorithm (dash-dotted line); analytical solution (dashed line). The noise in the adjoint MC is substantially less than that of the original algorithm at low frequencies. (b) Simulated (line) and measured (circles) thermal conductivities for pure nanocrystalline silicon with grain sizes at 550 nm, demonstrating good agreement. Inset: 2D schematic of a unit cell used in the simulation. The grain boundary bisects the domain at the indicated location.} 
\label{fig:Validation}
\end{figure*}

We validated our code by computing the spectral thermal conductivity versus phonon frequency of the bulk silicon at room temperature. The analytical expression for the spectral thermal conductivity is $k(\omega,p) = \frac{1}{3}C(\omega,p)v_g(\omega,p)\Lambda(\omega,p)$, where $C(\omega,p) = \hbar\omega D(\omega,p)\frac{\partial f_{BE}}{\partial T}$ is the mode specific heat and $\Lambda(\omega,p) = \tau(\omega,p)v_g(\omega,p)$ is the phonon MFP. The plot gives a quantitative measure how each phonon contributes to the total thermal conductivity. As shown in Fig.~\ref{fig:Validation}(a), excellent agreement between the computation and the analytical solution is observed. Also note that the stochastic noise is low everywhere over the entire phonon spectrum in the adjoint method, while the noise increases as the phonon frequency decreases with the original algorithm due to undersampling of low frequency phonons. 

With this validation, we proceed to the nanocrystalline material by determining the fitting parameter $\gamma$ in Eq.~(\ref{Eq:Transmissivity}). The reported material had 550 nm grains and was nominally undoped. By fitting the experimental thermal conductivity with our model, we obtain $\gamma = 0.12$ and the surface roughness $\eta = 0.5$ nm in Eq.~(\ref{Eq:specularity}) explain the thermal conductivity data from Wang \emph{et al.} Since the reported Si is nominally undoped, we take $A$ in Eq.~(\ref{Eq:Tau_md}) to be the value of bulk silicon. As shown in Fig.~\ref{fig:Validation}(b), the simulated thermal conductivities of nanocrystalline silicon obtained using the proposed transmissivity model is in good agreement with the experiment over the temperature range. Note that based on the fitted $\gamma$, the transmissivity varies from 1 to 0.85 as frequency increases, which is much higher than the typical nominal value of the transmissivity used in the literature.

\begin{figure*}[t!]
\centering
\includegraphics[scale=0.33]{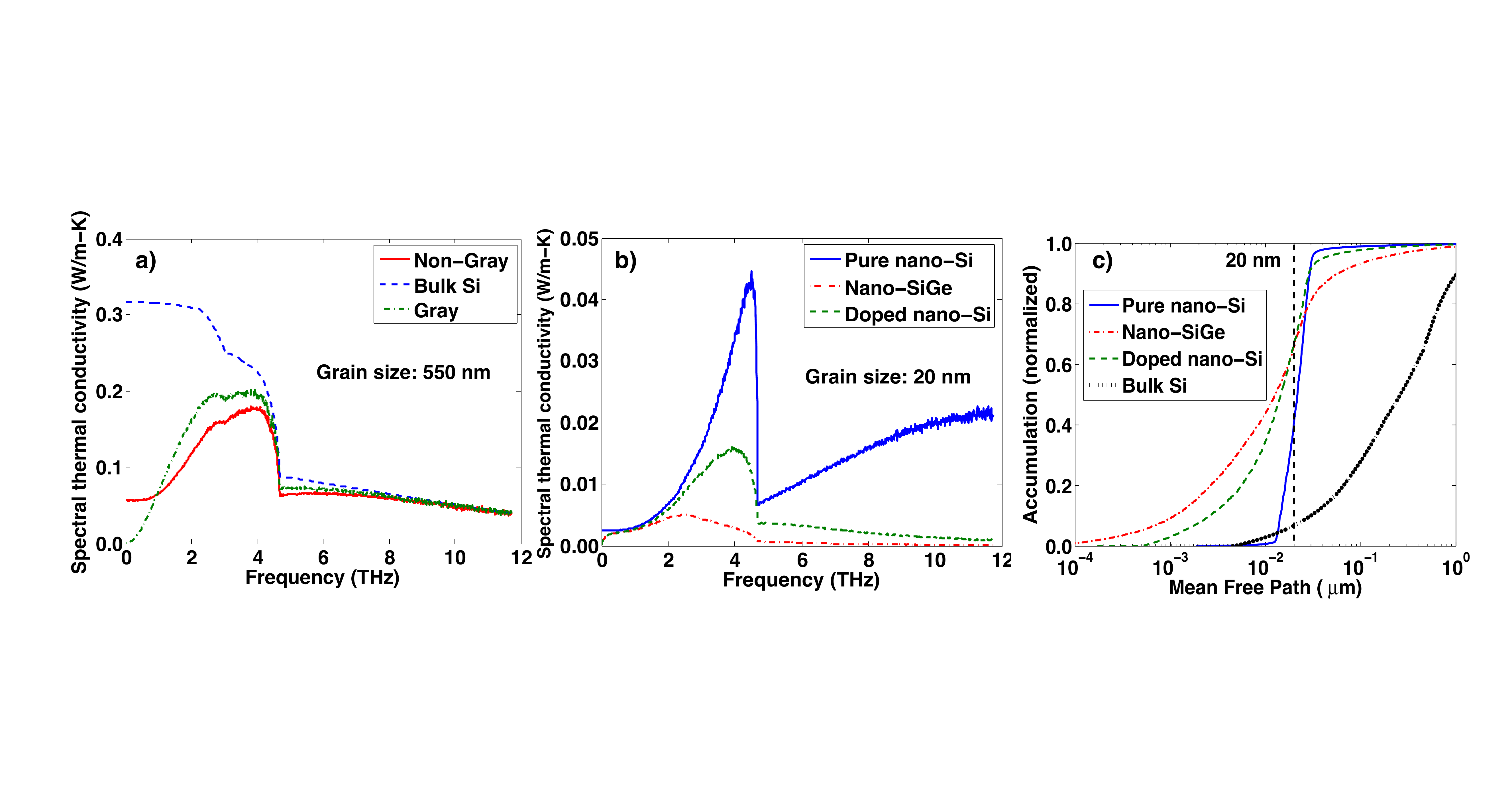}
\caption{(a) Room-temperature spectral thermal conductivity versus phonon frequency with 550 nm grain: bulk silicon (dashed line), nanocrystalline silicon with gray model (dash-dotted line) and non-gray model (solid line). Low frequency phonons carry more heat in the non-gray model. (b) Room-temperature spectral thermal conductivity versus phonon frequency with 20 nm grain: pure nanocrystalline silicon (solid line), doped nanocrystalline silicon (dashed line) and nanocrystalline silicon germanium (dash-dotted line). High frequency phonons are scattered by mass defects while low frequency phonons are scattered by electrons. (c) MFP distribution versus phonon MFP: pure nanocrystalline silicon (solid line), doped nanocrystalline silicon (dashed line), doped nanocrystalline silicon germanium (dash-dotted line) and bulk silicon (dotted line). Heat is still carried by long MFP phonons in nanocrystalline Si/SiGe even though many low frequency phonons are scattered by electrons. }
\label{fig:NanoSi}
\end{figure*}

With the fitting parameters determined, we first examine the difference between the gray and non-gray model for the same undoped nano-Si material. We focus on examining the distribution of heat in the thermal phonon spectrum by plotting the spectral thermal conductivity as a function of phonon frequency. We choose the frequency-independent transmissivity in the gray model to be 0.9 such that the total thermal conductivities given by the gray and non-gray models are approximately the same. As shown in Fig.~\ref{fig:NanoSi}(a), the gray model predicts a substantial decrease in heat at the low frequency regime while the non-gray model predicts a much higher contribution to the total heat from those low frequency phonons. This increase is expected because the transmissivity approaches one in the low frequency limit, indicating that low frequency phonons are scattered less by grain boundaries compared to the prediction of the gray model.

We next seek to understand doped nanocrystalline Si/SiGe thermoelectric materials by using the fitted constants described in the previous section to account for the mass defect and electron-phonon scattering mechanisms. We take grain size to be 20 nm following the experimental reports.\cite{Bux:2009,SiGeNanoLett,XWWang2008,Ren2009} The result of these effects on the heat transport in nanocrystalline silicon and SiGe is shown in Fig.~\ref{fig:NanoSi}(b). We find that high frequency phonons are strongly scattered by the impurities while low frequency phonons are scattered by the electrons, as expected. However, due to the non-gray transmissivity, a large fraction of heat is carried by phonons with frequency less than 4 THz. This observation is unexpected because under the gray model, thermal conductivity reduction is largely due to the scattering of these low frequency phonons, but contribution of these modes remains significant in the non-gray model. 

To determine the key length scales for heat conduction, we calculate the phonon MFPs. We plot this quantity as the MFP accumulation function, which has been shown to be a useful quantity for understanding thermal transport.\cite{damesCRC,Dames2013} We calculate the accumulation function by determining an effective MFP for each mode that incorporates all the scattering mechanisms, including grain boundary scattering, using the spectral thermal conductivity for each phonon frequency and polarization and the kinetic equation: $\Lambda_{eff} = 3 k(\omega,p)/C(\omega,p)v(\omega,p)$. The spectral thermal conductivity can then be sorted by MFP, from which the MFP accumulation is obtained from the cumulative sum of the spectral thermal conductivity.

Our simulations show a surprising result. As shown in Fig.~\ref{fig:NanoSi}(c), at room temperature, about 60 \% of the total heat in undoped nanocrystalline silicon with a grain size around 20 nm is carried by the phonons with mean free paths longer than the grain size. Even in doped nanocrystalline Si and SiGe, for which low frequency phonons are scattered by electrons, as much as 35 \% of the total heat is carried by these long MFP phonons.

To verify that our conclusion does not depend on the fitting parameters for mass-defect and electron-phonon scattering, we examine the change in the MFP distribution as $\tau^{-1}_{MD}$ and $\tau^{-1}_{ep}$ varied by 50\% to 200\% of their nominal values for nanocrystalline SiGe at room temperature. We find that there is only 3\% variation in heat contribution from long MFP phonons due to the change in mass defect scattering magnitude. Similarly, 7 \% variation is observation when varying $\tau^{-1}_{ep}$ using the same method. Thus, our observation of the importance of long MFP phonons is not sensitive to the assumptions made for these scattering mechanisms. 

\begin{figure*}[t!]
\centering
\includegraphics[scale=0.33]{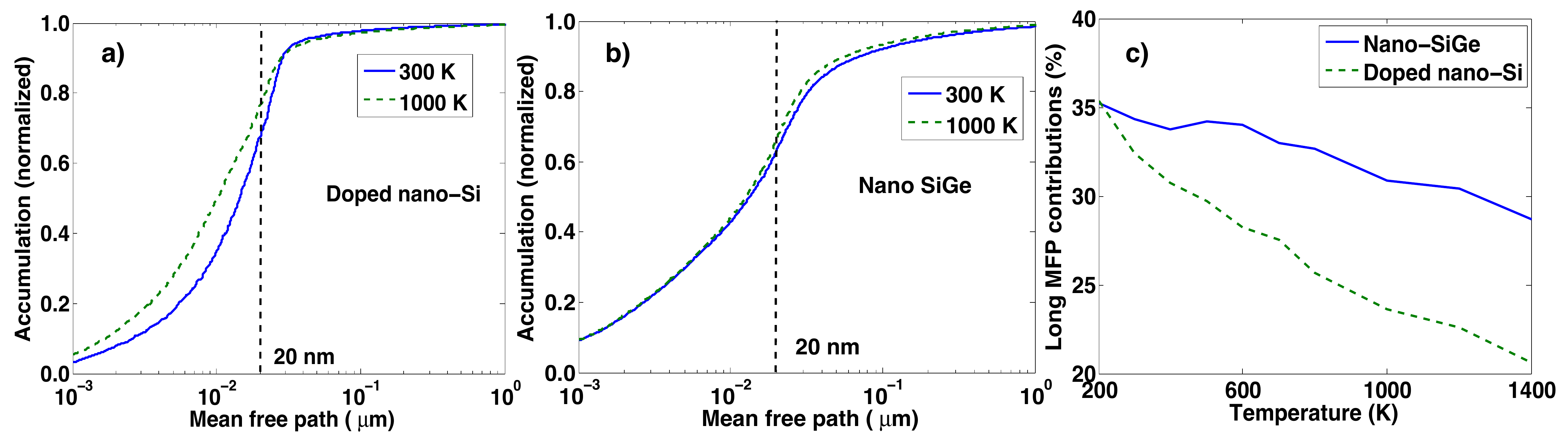}
\caption{MFP accumulation for (a) doped nanocrystalline Si and (b) doped nanocrystalline SiGe with 20 nm grain versus phonon MFP at 300 K (solid line) and 1000 K (dashed line). At high temperatures, there is still a large fraction of heat carried by long MFP phonons even when phonon-phonon scattering is dominant. (c) Percentage contribution from phonons with MFPs longer than the grain size versus temperature in doped nanocrystalline SiGe (solid line) and doped nanocrystalline Si (dashed line). The percentage contribution decreases as temperature increases but is still considerable at high temperatures.}
\label{fig:TempDep}
\end{figure*}

The effects of temperature on the MFP accumulations for doped nanocrystalline Si/SiGe are shown in Fig.~\ref{fig:TempDep}. As the temperature increases, the contribution from long MFP phonons decreases due to the dominance of phonon-phonon scattering. For doped nano-Si (Fig.~\ref{fig:TempDep}(b) \& dashed line in Fig.~\ref{fig:TempDep}(c)), at room temperature, about 35 \% of the total heat is due to phonons with MFPs longer than the grain size while at 1000 K, corresponding to the typical operating temperature of devices that use Si/SiGe, the percentage contribution decreases to 20 \%. On the other hand, this dominance shift is less obvious in nanocrystalline SiGe since most of the high frequency phonons , which are more likely to undergo phonon-phonon scattering, are already strongly scattered by the alloy atoms, as shown in Fig.~\ref{fig:TempDep}(a). At 1000 K, the contribution from these long MFP phonons in nano SiGe is still as high as 30 \% as shown by solid line in Fig.~\ref{fig:TempDep}(c). 

\begin{figure*}[t!]
\centering
\includegraphics[scale=0.5]{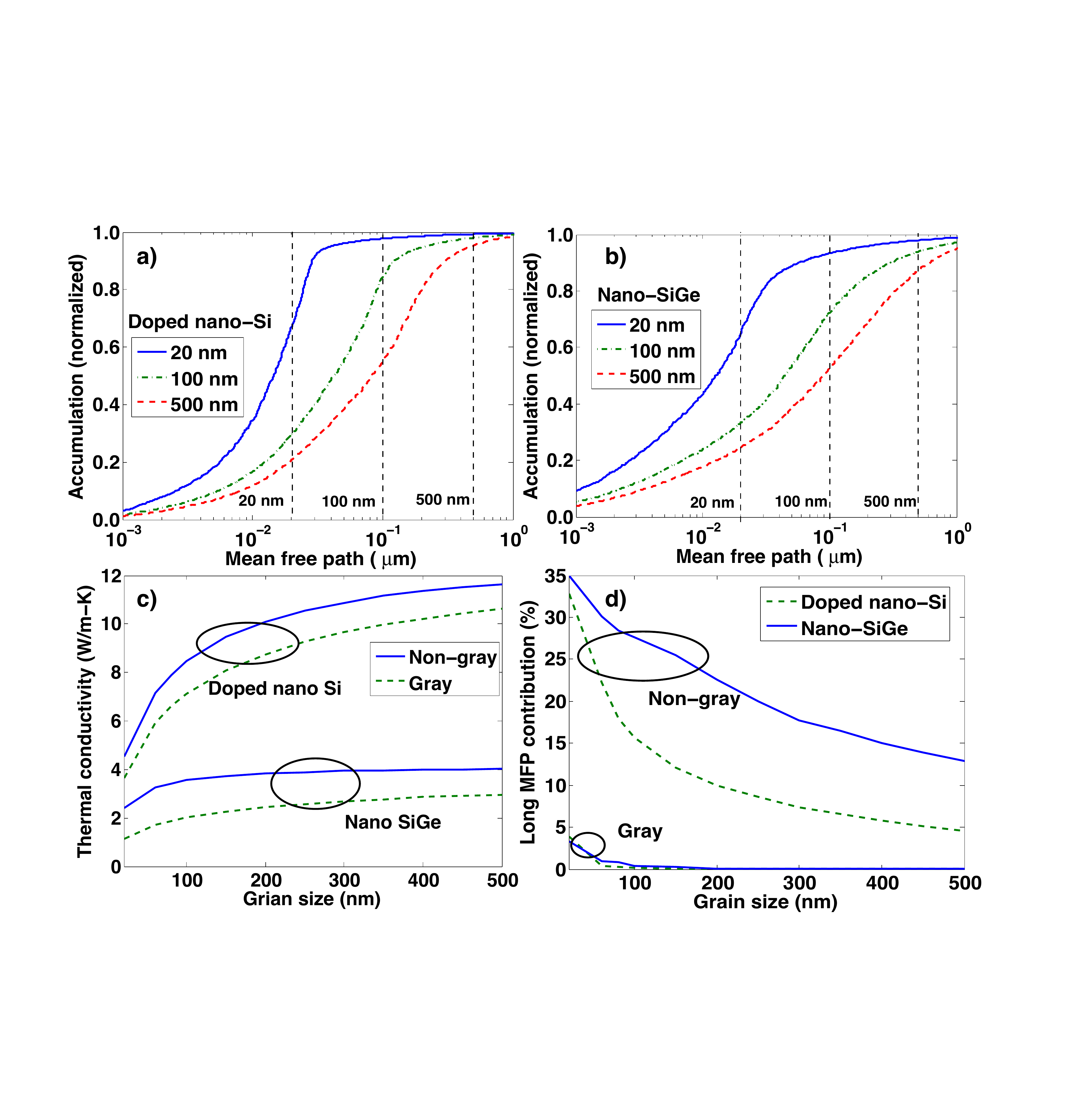}
\caption{MFP accumulation for (a) doped nanocrystalline Si and (b) doped nanocrystalline SiGe at 300 K versus phonon MFP with grain sizes at 20 nm (solid line), 100 nm (dash-dotted line) and 500 nm (dashed line). As the grain size decreases, the distribution shifts to shorter MFPs but the heat contribution from long MFP phonons increases. (c) Thermal conductivity versus grain size in doped nanocrystalline Si/SiGe using non-gray (solid lines) and gray (dashed lines) models. Thermal conductivity decreases as the grain size decreases, particularly for grains smaller than 100 nm. (d) Percentage contribution from phonons with MFPs longer than the grain size versus grain size in doped nanocrystalline Si (dashed lines) and doped nanocrystalline SiGe (solid lines) using non-gray and gray models. The contribution from the long MFP phonons increases as grain size decreases.}
\label{fig:GrainDep}
\end{figure*}

We now examine the effects of grain size on the heat transport in doped nanocrystalline Si/SiGe. Figures~\ref{fig:GrainDep}(a) and (b) show the MFP accumulation versus phonon MFP for doped nanocrystalline Si/SiGe with different grain sizes at 300 K. Since the mass defect scattering strength is different in the two materials, short MFP phonons are less scattered by the mass defects and in general carry more heat in doped nanocrystalline Si than in doped nanocrystalline SiGe. Regardless of the strength of the mass defect scattering, we find an increase in contribution from long MFP phonons as the grain size decreases in both materials even though smaller grain sizes result in shorter MFP phonons.

To clearly demonstrate this trend, we plot both thermal conductivity and long MFP contribution versus grain size in Figs.~\ref{fig:GrainDep}(c) \& (d). Previous studies have shown that decreasing the grain size can effectively reduce the thermal conductivity.\cite{Cruz:2013,Colombo:2014} The decreasing trend of thermal conductivity as grain size decreases in Fig.~\ref{fig:GrainDep}(c) confirm this observation. As the grain size varies from 550 nm to 20 nm, there is a 30 \% reduction of thermal conductivity for nanocrystalline SiGe and 45 \% in doped nanocrystalline Si. Due to the stronger mass defect scattering, the thermal conductivity of nanocrystalline SiGe is more than 50 \% lower than that of doped nanocrystalline Si at a given grain size. When we look into the contribution to the total thermal conductivity from each phonon mode, we find that, counterintuitively, the contribution from those phonons with MFPs longer than the grain size increases as the grain size decreases as shown in Fig.~\ref{fig:GrainDep}(d). For a 20 nm grain, the contribution from long MFP phonons is as high as 35 \% in both nanocrystalline Si and SiGe. This result can be explained by the increasing fraction of phonons with MFP that is longer than grain size as the grain size decreases. According to the frequency-dependent transmissivity, those phonons are less affected by the grain boundaries compared to the prediction of the gray model. Therefore, their contribution to the total thermal conductivity increases even though the total thermal conductivity decreases. 

These observations indicate that nanocrystalline grain boundaries may not be as effective as previously believed at scattering long MFP phonons. If a grain boundary can be designed such that it diffusely scatters all phonons independent of frequency as in the gray model, an additional reduction in the phonon thermal conductivity can be achieved. In Figs.~\ref{fig:GrainDep}(c) \& (d), we plot the thermal conductivity and long MFP phonon contribution versus grain size for the gray model with a constant transmissivity of 0.85, corresponding to the smallest value of the transmissivity in the non-gray model. Figure~\ref{fig:GrainDep}(c) shows that thermal conductivity is approximately 12 \% lower than non-gray model in doped nanocrystalline silicon and 30 \% lower in doped nanocrystalline SiGe. The higher reduction in SiGe is because long MFP phonons in doped nanocrystalline SiGe, which are strongly scattered by the grain boundary, contribute more to thermal conductivity than in doped nanocrystalline Si. Figure~\ref{fig:GrainDep}(d) quantifies this reduction in contribution from long MFP phonons. The gray grain boundary scatters most of the long MFP phonons such that their contribution to the total thermal conductivity is negligible. In both doped nanocrystalline Si and SiGe with 20 nm grains, the long MFP contribution is reduced to less than 5 \%. Further thermal conductivity reduction can be achieved if the transmissivity is further decreased. Therefore, great potential to further increase the thermoelectric performance of nanocrystalline Si/SiGe exists if such grain boundaries can be designed.  

\section{Conclusions}

Nanocrystalline Si and SiGe have great potential as thermoelectrics, but a better understanding of grain boundary scattering is necessary to improve their efficiency. In this work, we have studied thermal phonon transport in the full 3D geometry of nanocrystalline Si and SiGe using a novel adjoint variance-reduced energy-based Monte Carlo method. We find that low frequency, long MFP phonons, which are previously predicted to carry negligible heat in the gray model, may still carry a substantial amount of heat due to the frequency-dependent grain boundary scattering. Significant potential to improve the efficiency of nanocrystalline Si and SiGe exists if these phonons can be scattered. Our work provides important insight into how to further increase the thermoelectric performance of nanostructured silicon and silicon-germanium alloys.

\section*{Acknowledgements}

The authors thank J.P. Peraud and N.G. Hadjicontantinou for useful discussions. This work was sponsored in part by Robert Bosch LLC through Bosch Energy Research Network Grant no. 13.01.CC11, by the National Science Foundation under Grant no. CBET 1254213, and by Boeing under the Boeing-Caltech Strategic Research \& Development Relationship Agreement.

\end{document}